\documentclass[a4paper,11pt]{article}
\usepackage{pos}
\usepackage{graphicx}
\usepackage{float} 
\usepackage{subfigure}
\usepackage{url}

\title{Research and progress of front-end readout prototype system for GRANDProto300}
 \ShortTitle{Front-end readout system for GRANDProto300}

\author*[a,b]{Xing Xu}
\author[a,b]{Jianhua Guo}
\author[a]{Shen Wang}
\onbehalf{for the GRAND Collaboration}

\affiliation[a]{Purple Mountain Observatory, Chinese Academy of Sciences,\\
No.10, Yuanhua Road Qixia District, Nanjing, China}

\affiliation[b]{University of Science and Technology of China, School of Astronomy and Space Science,\\
No.96, JinZhai Road Baohe District, Hefei, China}

\emailAdd{xingxu@pmo.ac.cn}

\abstract{GRANDProto300 is the planned 300-antenna pathfinder array of the Giant Radio Array for Neutrino Detection (GRAND), of which the first 100 detection units have been already produced. Its main goal is to demonstrate the viability of the detection of the radio emission from air showers initiated by inclined ultra-high-energy cosmic rays with energies of $10^{16.5}$ to $10^{18.5}$ eV, covering the purported transition region from their Galactic to extragalactic origin. The front-end readout system of each detection unit of GRANDProto300 processes signals from the radio antenna and the particle detector, generates the first-level trigger, and communicates with a central processing station. Based on earlier designs, we have built the first prototype of this system using two development boards and one self-designed front-end board. We present our new design that is improved and more economical than the earlier one, as well as test results and prospects for future work.}

\ConferenceLogo{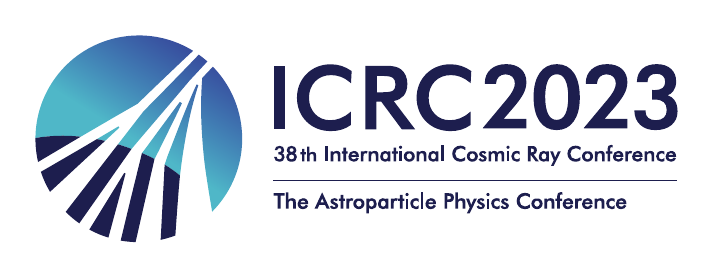}

\FullConference{%
38th International Cosmic Ray Conference (ICRC2023)\\
  26 July - 3 August, 2023\\
  Nagoya, Japan}

\begin{document}
\maketitle

\section{Introduction}
\begin{figure}[b]
    \centering
    \includegraphics[scale=0.39]{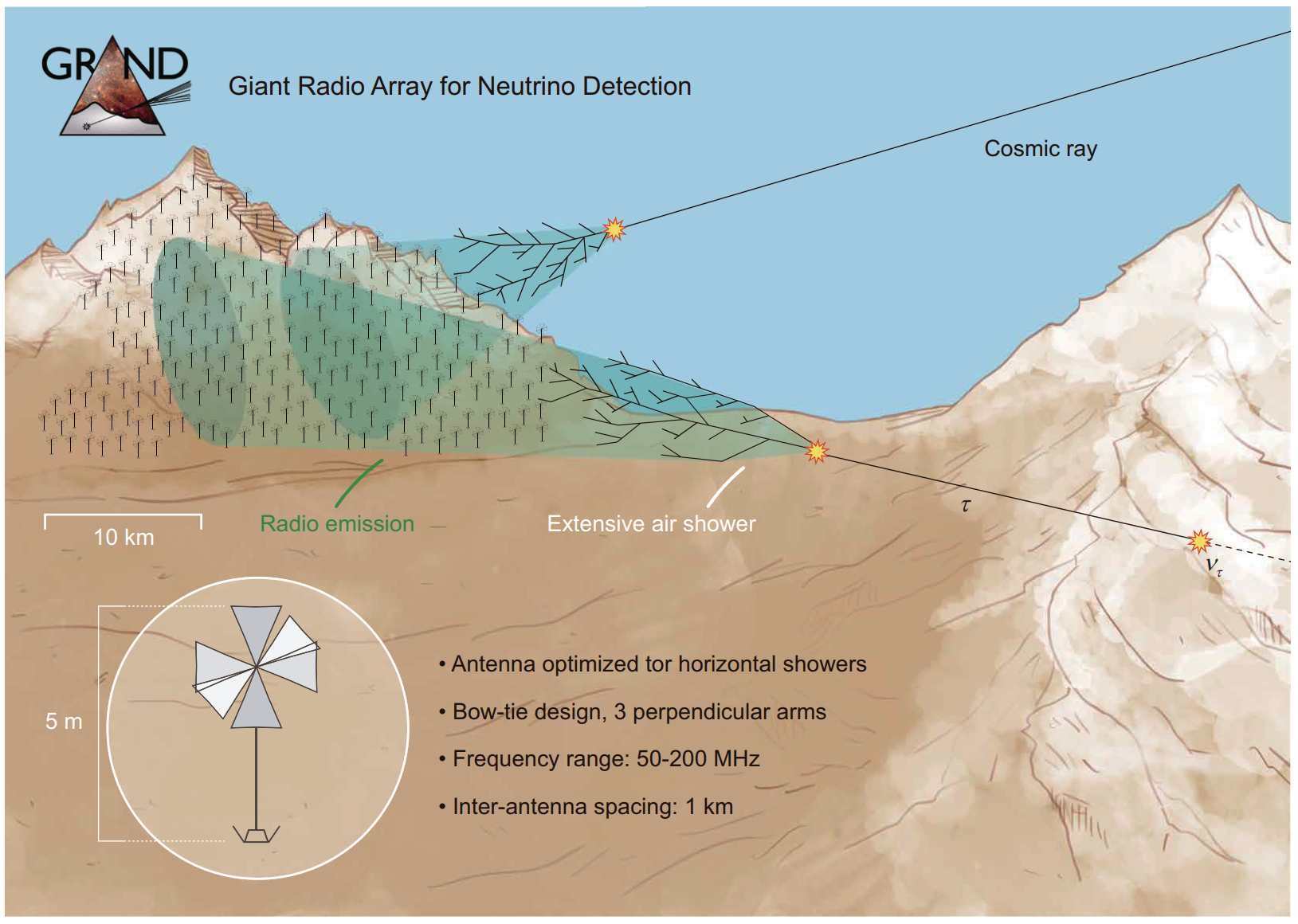}
    \caption{GRAND detection strategy \cite{GRAND:2018iaj} }
    \label{Fig.1}
\end{figure}
After more than twenty years of development, the radio-detection of cosmic rays and neutrinos has become a feasible detection strategy, one that can complement and extend traditional methods of particle detection \cite{Schroder:2016hrv}. Its main advantages are low cost, nearly all-weather operation, and higher sensitivity to very inclined extensive air showers, such as the ones expected to be triggered by ultra-high-energy (UHE) particles in the atmosphere. With the development of antennas and electronics and deeper understanding of radio emission from high-energy particle showers, the upcoming experiments that target the radio-detection of cosmic rays or neutrinos are moving from external trigger to self-triggering operation, and expanding the construction from small arrays to large arrays.

The Giant Radio Array for Neutrino Detection (GRAND) \cite{GRAND:2018iaj} uses radio methods to detect high-energy particles in the frequency band of 50--200 MHz, as shown in Figure \ref{Fig.1}. What we already know is that energetic particles produce showers in the atmosphere and radio signals under the action of the geomagnetic field effect and the Askaryan effect. If the radio signals are captured and converted into analyzable data, we can reconstruct information such as the energy and mass composition of particles. Besides detecting UHE neutrinos above $10^{17}$ eV, GRAND will be a large detector of ultra-high-energy cosmic rays (UHECRs). It is expected to give revolutionary insights into the origin of UHECRs and the nature of UHE neutrino sources.

GRAND expects to eventually build a giant array containing 200,000 antennas, distributed in multiple, separate sub-arrays, through a phased construction plan, and the project is currently in the GRANDProto300 (GP300) phase, which is a 300-antenna pathfinder array. Its main goal is to demonstrate the viability of the detection of the radio emission from air showers initiated by inclined UHECRs with energies of $10^{16.5}$ to $10^{18.5}$ eV, covering the purported transition region from their Galactic to extragalactic origin.

\section{GP300 detection unit}
As a self-triggering radio detection array, radio background noise level is an important criterion of site selection. Considering the ease of access, infrastructure, support by local authorities, and possible extension to larger arrays, the selection of experimental sites can be challenging. Since August 2017, the project team has visited more than 10 candidate sites in western China, and has selected a site near Dunhuang City, Gansu Province. Thirteen sets of detection units and their corresponding central processing stations have been deployed in this site. Equipment commissioning and trial operation are underway, and the system is currently running well \cite{pos_mpx}.

\begin{figure}[h]
\centering
\subfigure{
\label{Fig.2.1}
\includegraphics[scale=0.18]{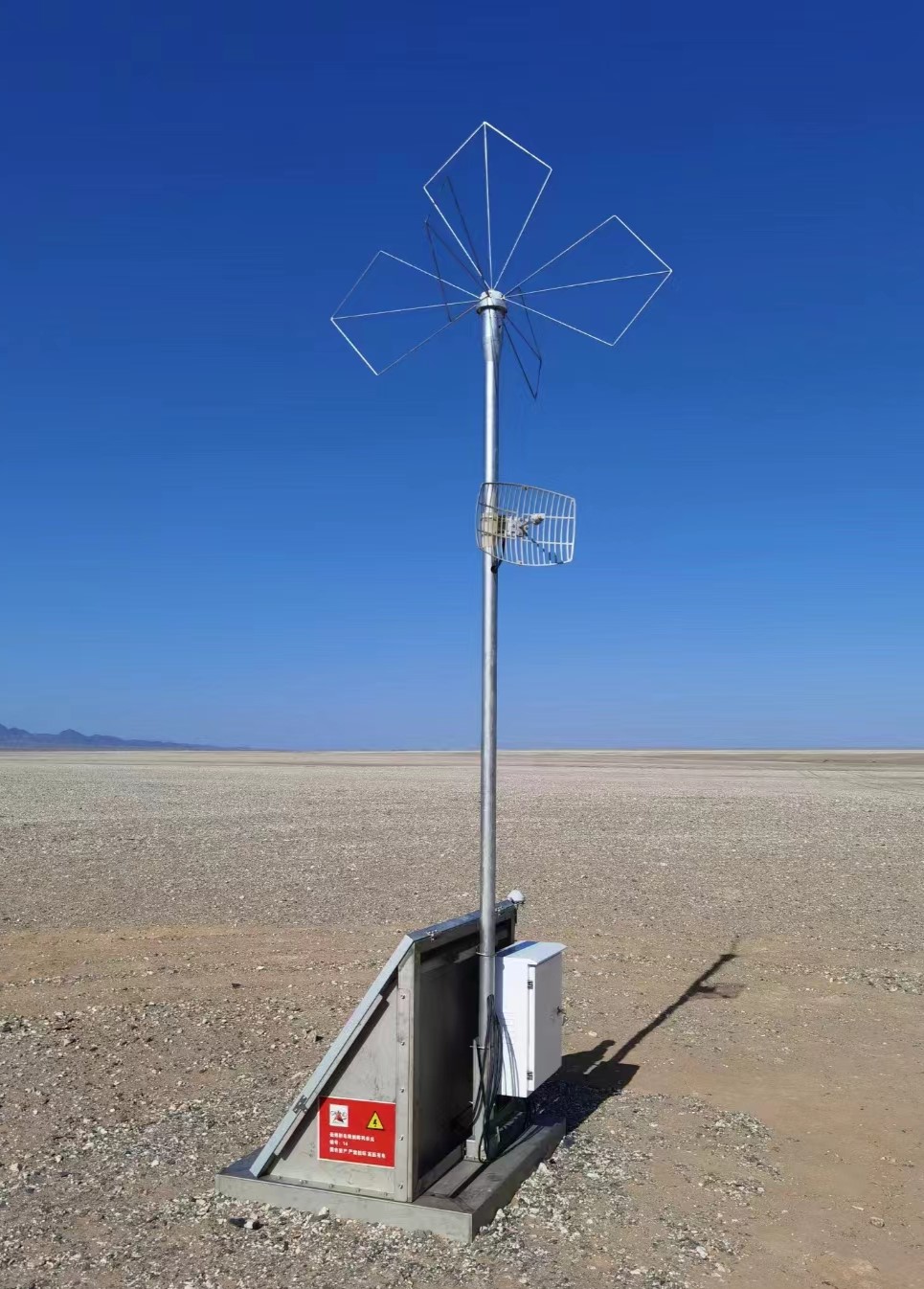}
}
\subfigure{
\label{Fig.2.2}
\includegraphics[scale=0.62]{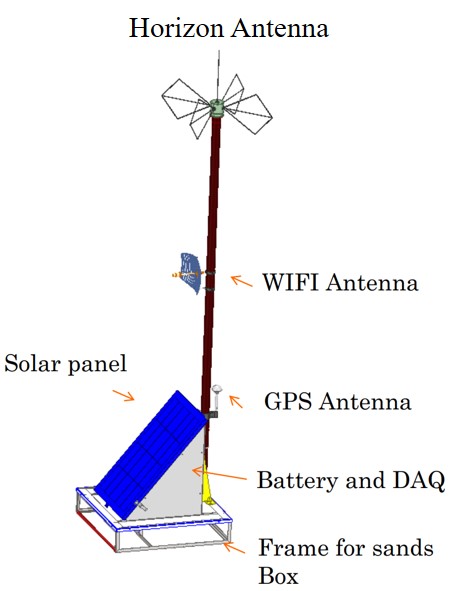}
}
\caption{A picture of the detection unit at the Dunhuang site (left) and its schematic view (right). The DAQ board and the battery with charge controller are placed in the boxes.}
\label{Fig.2}
\end{figure}

Figure \ref{Fig.2} shows one GP300 detection unit. A detection unit of GP300 is composed of a radio antenna, WiFi antenna, GPS receiver module, solar panel, data acquisition (DAQ) system and mechanical structure. The antenna is a 3-arm butterfly-type antenna placed 3 meters above the ground. The solar panel and the battery (not shown in the figure) grant the detection unit an independent supply of power.

At the GP300 prototype stage, we also plan to add particle detectors to additionally detect the air shower particles. Thanks to the GPS information, the results of the individually triggered particle detectors can be compared with the radio channels, resulting in a more accurate reconstruction of the detected showers.

\section{The front-end readout system}
When it comes to the consideration of the front-end readout scheme, there are mainly two technical routes. The first one is adopted in experiments such as LOFAR \cite{Nelles:2014dja} and AERA \cite{Kelley:2012agy}. The voltage signal generated in the radio antenna is firstly amplified and filtered, then continuously sampled and quantified by an analog-to-digital converter (ADC), the digital signal is processed in chips like the field-programmable gate array (FPGA) in the end. Another route is used in experiments such as ARIANNA \cite{Barwick:2014rca}. The voltage waveform is sampled at 1 or several Giga-samples-per-second and then stored in the switched capacitor array in an updated manner. If the system is triggered, the stored sampling points will be digitized by an ADC with a sampling rate of dozens of MHz, and then fed into the subsequent circuits. The advantage of the former scheme over the latter is that it is easy to implement at low frequencies and the technology is relatively mature, suitable for self-triggering systems.

The GRAND Collaboration has designed a complete front-end readout system named GRAND PROTO300 DIGITIZER V2 and produced 100 printed circuit boards. These boards are functional and are undergoing longer testing in the field. However, we identified two points necessitating further development. First, the high cost makes these boards unsuitable for the construction of subsequent large-scale arrays. Second, its reliability performance is not suitable for the extreme conditions of our current experimental site. In addition, there is room for improvement and upgrade of the system itself.

Therefore, we are designing and testing a new front-end readout prototype system, referred to as the prototype system in the following sections. This work is planned to be divided into two phases, firstly using two commercial development boards and one self-designed board for system construction. Then a complete prototype system will be constructed for verification. The following sections, unless otherwise noted, refer to the first phase of this work.

\subsection{Hardware}
\begin{figure}[b]
    \centering
    \subfigure{
    \label{Fig.3.1}
    \includegraphics[scale=0.6]{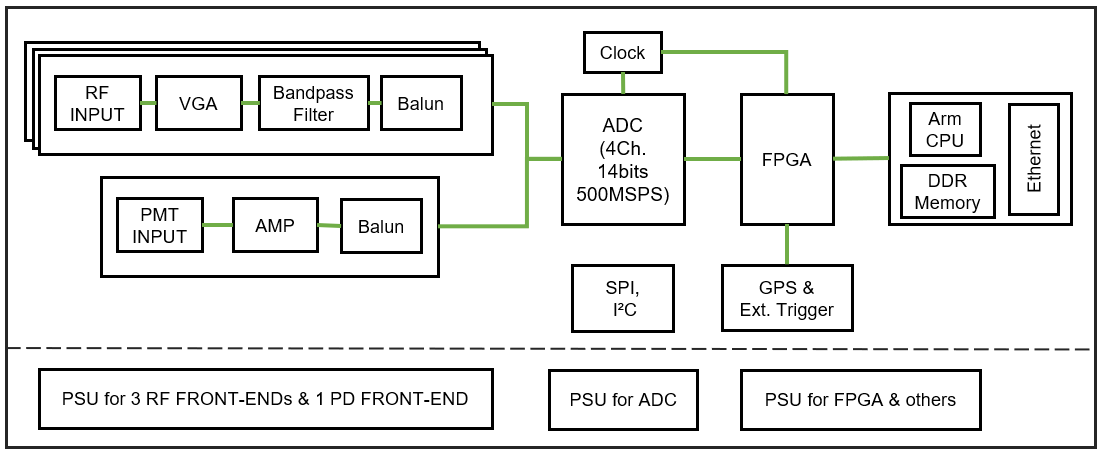}
    }
    \subfigure{
    \label{Fig.3.2}
    \includegraphics[scale=0.38]{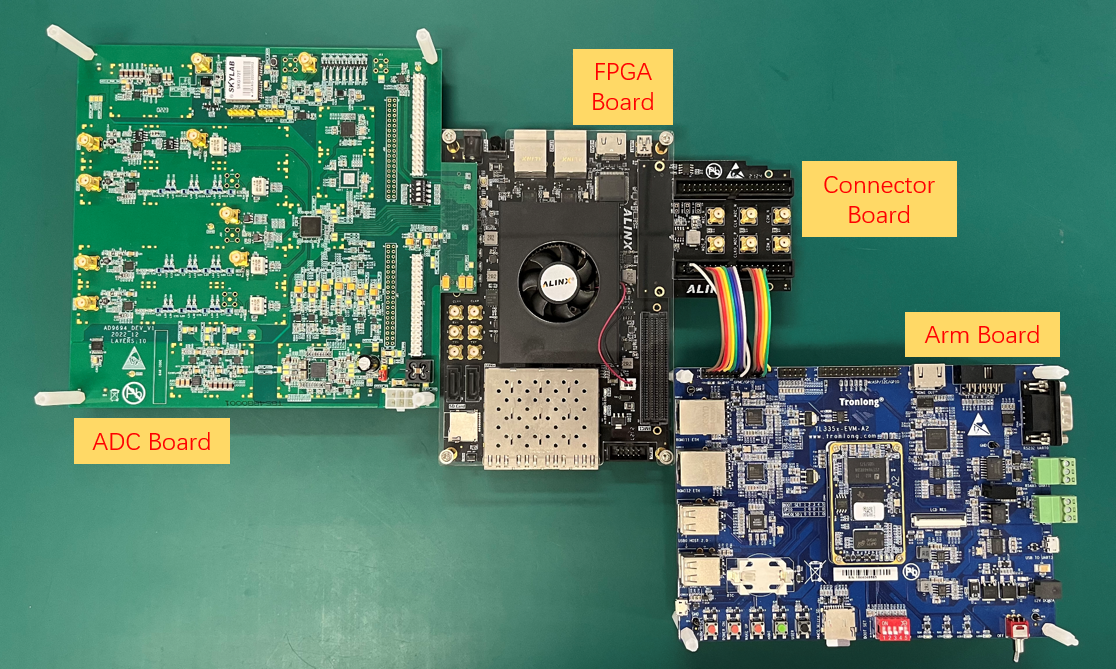}
    }
    \caption{Schematic (top) and picture (bottom) of the GP300 front-end readout prototype system}
    \label{Fig.3}
\end{figure}
Figure \ref{Fig.3} shows the hardware design of the prototype system. It is based on the GRAND PROTO300 DIGITIZER V2, but there are two main differences. First, the FPGA and the ARM CPU are separated, which greatly reduces the system cost without affecting the function and corresponding performance of the system. Second, a particle detector readout channel is added, using all channels of the ADC, improving the readout capability of the system.

The 3 radio antenna arms signals transmitted by the low-noise amplifier (LNA) and cable are re-amplified by the variable gain amplifier (VGA). They pass through a 30--210MHz band-pass filter circuit, together with another photomultiplier tube (PMT) signal from the particle detector, they are fed into a quad, 14-bit, 500 Msps AD9694 ADC for sampling and quantization. The resulting digital signal is transmitted via the FMC connector to the AKU040 for digital signal processing, which is a development board equipped with a XCKU040 FPGA chip from Xilinx UltraScale family. The FPGA and ARM CPU development board use an off-chip bus called General Purpose Memory Controller (GPMC) for instruction interaction and data transmission, the speed of which can reach at least 16 MBps. In the prototype system, the ARM CPU communicates with the test host through a Gigabit Ethernet port for transmission of data packages and a serial UART for command-line interaction.

\begin{figure}[h]
    \centering
    \includegraphics[scale=0.5]{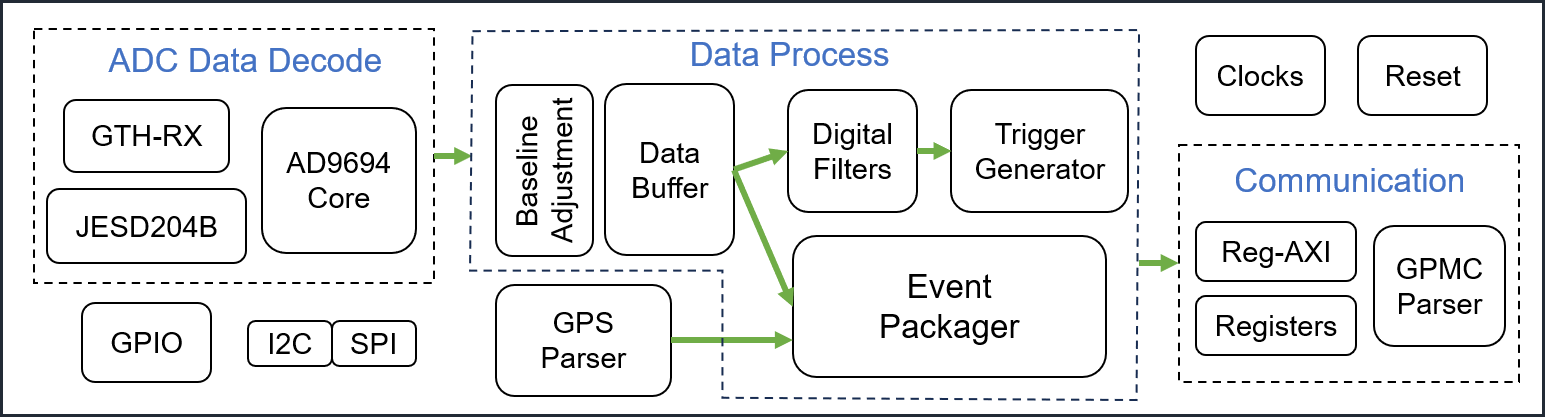}
    \caption{Diagram of the firmware logic of the GP300 front-end readout prototype system}
    \label{Fig.4}
\end{figure}

\subsection{Firmware and software}

Figure \ref{Fig.4} shows a framework diagram of the firmware logic of the prototype system running in the FPGA. The data from the 4 sampled channels of the 14-bits ADC are sent to the FPGA by the JESD204B protocol at a lane rate of 10 Gbps, they are parsed and then split into two branches after baseline adjustment. One is stored in the local memory, another is filtered and then passed through the trigger logic. If a local trigger is generated, a timestamp from the GPS receiver will be sent to the ARM CPU by the GPMC bus interface and then to the central DAQ by wireless transceiver. Furthermore, if the time and spatial coincidence of multiple detection units generates a high-level trigger in the central DAQ, the ADC raw data in each relational detection unit will be sent to the central DAQ and stored in the hard disk.

The baseline of each channel can be integrated over a selected time frame. This moving average will be subtracted from the ADC value before used in the trigger logic. To increase the signal-to-noise ratio by filtering out narrow-band transmitters from the digitized antenna signals, we use digital filters in the FPGA, such as a series of infinite-impulse response notch filters. A trigger algorithm with voltage thresholds and other parameters is used to capture radio pulses in the background containing man-made radio frequency interferences, details can be found in \cite{pos_pablo,pos_sandra}.

To test the prototype system, a simple test program is written using the C language. It can configure several chips on the ADC board, check and read event data from the FPGA. Furthermore, event packages will be parsed into readable text files which makes the test process easier.

\begin{figure}[t]
    \centering
    \includegraphics[scale=0.3]{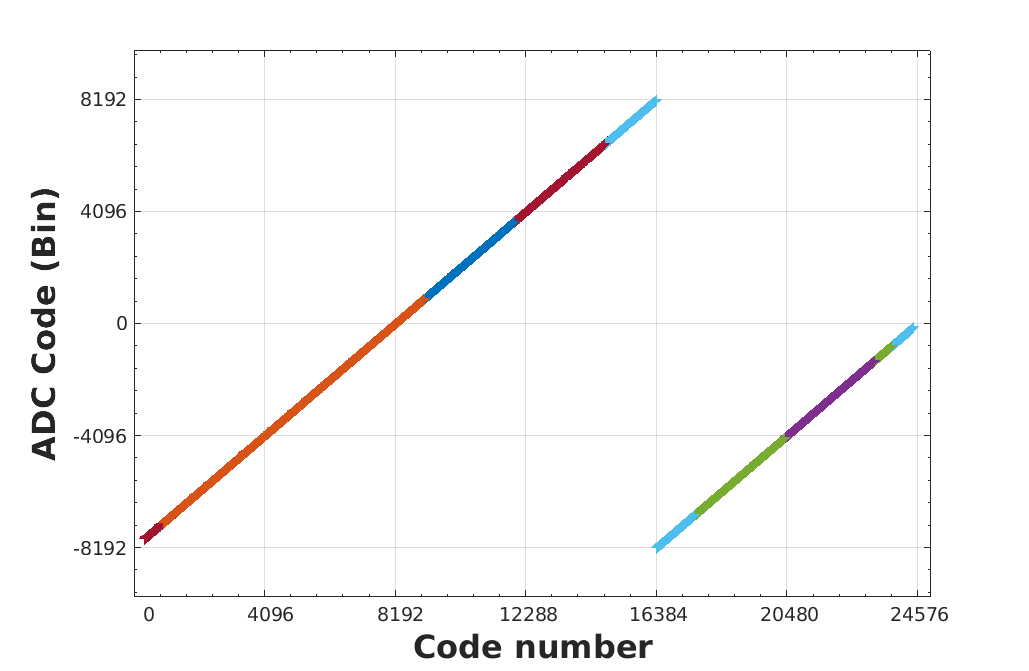}
    \caption{Test result of ADC code transmission. All 16384 bins are covered by 100 data packages, each of which contains 8192 sample points. 100$\%$ data transfer accuracy is achieved.}
    \label{Fig.5}
\end{figure}
\begin{figure}[t]
    \centering
    \includegraphics[scale=0.3]{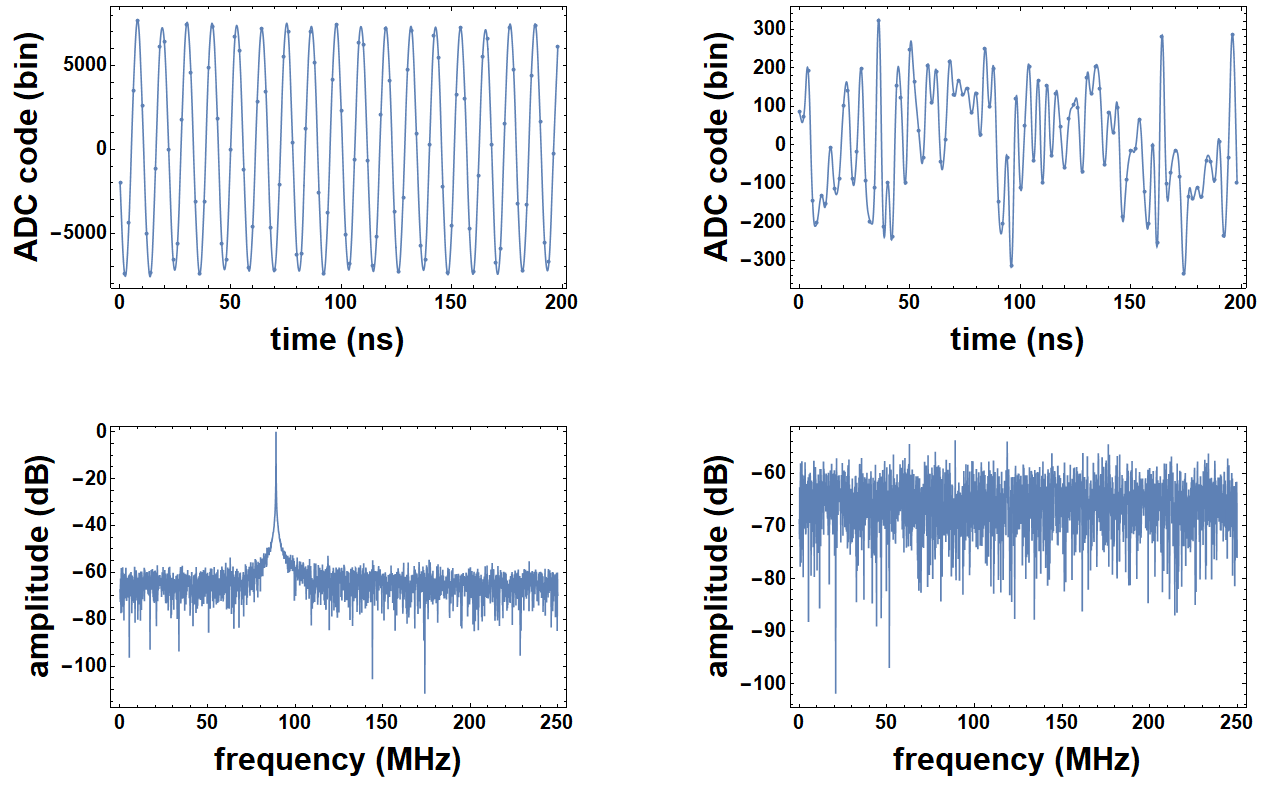}
    \caption{Test result of the notch filter. The target frequency signal is well filtered out of the background, as seen in the time domain of the raw trace (top left) and the filtered trace (top right), as well as in the frequency domain of the raw trace (bottom left) and the filtered trace (bottom right).}
    \label{Fig.6}
\end{figure}

\section{Tests}
After commissioning, we performed tests to verify the feasibility and performance of the prototype system. The test of the readout channel of particle detectors has not been completed, so the results about it are not shown here.

The first test regarded data acquisition and transmission. By setting the configuration registers inside the AD9694 ADC, we can make it work in one of its test modes, outputting ramp codes from -8192 to 8191. The data obtained in the ARM CPU using the 10-second trigger mode is shown in Figure \ref{Fig.5}. There is no error among 100 packages, each of which contains 8192 sample points, indicating that the converted codes of the AD9694 can be transmitted correctly through the entire system.

The second test regarded digital filtering and trigger logic. A trace of 89-MHz single-frequency sinusoidal wave mixed with Gaussian white noise is injected into the notch filter inside the FPGA. The results are shown in Figure \ref{Fig.6}. By comparing the results of running a fast Fourier transform on data points obtained with the notch filter bypassed and enabled, we find that the filter effectively reduces the amplitude of the input signal. Concerning the trigger logic test, we used a signal generator to generate simulated signals which can be triggered by the trigger algorithm theoretically and injected these signals into the ADC. The result showed that all injected signals are triggered by the system. It indicates that the trigger logic in the firmware works well, but further tests will be performed under field conditions.

We also tested the AD9694 chip which is one of the key units in this prototype system. The test methods are detailed in \cite{webpage1,webpage2}. Due to factors such as noise and distortion, the ADC cannot reach its number of bits in actual measurement. The effective number of bits (ENOB) indicates our ability to distinguish noise and signal from the ADC codes. It is tested to be 10.7 for AD9694 when the input frequency is 10.5 MHz, this result is within the range given by the chip datasheet \cite{webpage3}. Besides, we tested the differential and integral nonlinearities of the AD9694, the results are consistent with the test example results in the chip datasheet. We conclude that the AD9694 works as well as it could in our prototype system.

\section{Summary and outlook}

A custom circuit board and two development boards were used to build the first version of the new GRANDProto300 front-end readout prototype system. The tests showed that it works well on data collecting, digital filtering, and triggering. But more tests are needed to verify the feasibility of the readout of particle detectors.

Based on this, we have begun the hardware design of the second version of the prototype system. In addition to integrating the entire system on a single board, we also plan to add features such as low-power management design and remote firmware update, which can greatly improve the reliability of the system. In terms of firmware, new digital filters and trigger algorithms will be applied to assess their performance in the GRANDProto300 site in Dunhuang, China.

%
%
%

\end{document}